# *Ab initio* density functional theory study of uranium solubility in $Gd_2Zr_2O_7$ pyrochlore[*]

Chen Qing-yun(陈青云)[1,2] Shih Kai-min(施凯敏)[2;1)] Meng Chuan-min(孟川民)[3;2)] Liao Chang-zhong(廖长忠)[2] Wang Lie-lin(王烈林)[1] Xie Hua(谢华)[1] Lv Hui-yi(吕会议)[1] Wu Tao(吴涛)[1] Ji Shi-yin(吉诗银)[1] Huang Yu-zhu(黄禹竹)[1]

[1]*Key Subject Laboratory of National Defense for Radioactive Waste and Environmental Security, Southwest University of Science and Technology, Mianyang 621010, China*

[2]*Department of Civil Engineering, The University of Hong Kong, Pokfulam Road, Hong Kong SAR, China*

[3]*Key Lab for Shock Wave and Detonation Physics Research, Institute of Fluid Physics, Chinese Academy of Engineering Physics, Mianyang 621900, China*

In this study, an *ab initio* calculation is performed to investigate the uranium solubility in different sites of $Gd_2Zr_2O_7$ pyrochlore. The $Gd_2Zr_2O_7$ maintains its pyrochlore structure at low uranium dopant levels, and the lattice constants of $Gd_2(Zr_{2-y}U_y)O_7$ and $(Gd_{2-y}U_y)Zr_2O_7$ are generally expressed as being linearly related to the uranium content *y*. Uranium is found to be a preferable substitute for the B-site gadolinium atoms in cation-disordered $Gd_2Zr_2O_7$ (where gadolinium and zirconium atoms are swapped) over the A-site gadolinium atoms in ordered $Gd_2Zr_2O_7$ due to the lower total energy of $(Gd_{2-y}Zr_y)(Zr_{2-y}U_y)O_7$. The theoretical findings present a reasonable explanation of recent experiment results.

Submitted to 'Chinese Physics C'

*Supported by National Natural Science Foundation of China (11304254), General Research Fund Scheme of the Research Grants Council of Hong Kong (715612, 17206714), and Key Subject Laboratory Foundation of National Defense for Radioactive Wast Environmental Security of SWUST (13zxnk09, 13zxnk11)

1）E-mail: kshih@hku.hk

2）E-mail: mcm901570@126.com

# 1 Introduction

The emergence of nuclear energy offers a promising opportunity for developing low-cost and highly efficient energy sources. However, the safe disposal of high levels of nuclear waste, especially long-lived radioactive elements, remains an important challenge in the nuclear industry [1,2]. Gadolinium zirconate ($Gd_2Zr_2O_7$) pyrochlore exhibits a high chemical durability, radiation resistance and solubility for large radionuclide species (such as thorium, uranium and plutonium) and is therefore an attractive candidate for nuclear-waste host material [3-8].

In an ordered $A_2B_2O_7$ pyrochlore structure (space group Fd3m) with A (gadolinium or rare earths) and B (zirconium, titanium, tin, hafnium or plumbum) cations ordered on the *16d* (0.5, 0.5, 0.5) and *16c* (0, 0, 0) sites, respectively, oxygen can be found at the *48f* ($x$, 0.125, 0.125) and *8b* (0.375, 0.375, 0.375) positions. The *8a* (0.125, 0.125, 0.125) is a vacant site (using the Wyckoff notation)[9-10]. $Gd_2Zr_2O_7$ crystallizes in the cubic pyrochlore structure, and large amounts of actinide elements are expected to be incorporated into the matrix at both the gadolinium and zirconium lattice positions [11-14].

Sickafus *et* al. [6] concluded that radiation-tolerant materials in the 2:2:7 pyrochlore stoichiometry phase present no advantage. The key to radiation tolerance is an inherent ability to accommodate atomic lattice disorder. Under ion irradiation, the ordered pyrochlore superstructure transforms into an anion-disordered pyrochlore before its final transformation into a cation-disordered defect-fluorite structure type [15]. Studies have shown that $Gd_2Zr_2O_7$ transforms into a radiation-resistant anion-deficient fluorite structure upon irradiation and shows more radiation resistance than $Gd_2Ti_2O_7$, which has a higher defect formation energy disorder and is more susceptible to amorphization [3].

Different experiments have arrived at inconsistent conclusions related to the primacy of cation or anion disorders. Diffraction studies have indicated that oxygen disorder causes cation disorder, and spectroscopic studies have yielded the opposite conclusion [10]. Although the underlying mechanisms of defect formation and radiation resistance are not yet well understood, there is a consensus that the accommodation of lattice disorder should improve amorphization resistance in a displacive radiation damage environment [3, 6, 9].

Kulkarni *et* al. [16] reported that plutonium can completely replace gadolinium (A site) in a $Gd_2Zr_2O_7$ matrix. However, although uranium can replace only 40% of the gadolinium at the A site in the $Gd_2Zr_2O_7$ matrix, it can replace all of the zirconium at the B site. The structure of $Gd_2Zr_2O_7$ transforms from pyrochlore into a closely related fluorite structure as the level of uranium dopant increases [11]. By controlling the sintering atmosphere, uranium with different oxidation states and ionic radii can be incorporated in a $Gd_2Zr_2O_7$ matrix along with pyrochlore and defect fluorite structures [17].

Gregg *et a*l. [18] recently examined U-doped and off-stoichiometric $Gd_2Zr_2O_7$ pyrochlore via X-ray diffraction (XRD) and X-ray absorption near-edge structure (XANES) spectroscopy. They suggested that uranium cations are largely located at pyrochlore B sites instead of the targeted A sites. The study provides direct evidence for cation antisite disorder in U-doped $Gd_2Zr_2O_7$ pyrochlore [18]. Although some experimental studies have considered the solution of high-level radioactive-waste uranium in $Gd_2Zr_2O_7$ pyrochlores, the mechanisms underlying uranium doping and disorder (defect formation) in $Gd_2Zr_2O_7$ crystals are not yet well understood.

*Ab initio* density functional theory (DFT) simulations have become ideal tools for systematically investigating properties rather than chemical compositions. In this study, the virtual crystal approximation (VCA) method is used to calculate the total energies and structure properties of U-doped $Gd_2Zr_2O_7$ pyrochlores (ordered and cation-disordered structures). The main purpose of this study is to clarify the solution behavior of uranium in $Gd_2Zr_2O_7$ pyrochlores.

## 2 Computational details

The calculations are performed using the first-principles plane-wave pseudopotential method based on DFT incorporated into the *CASTEP* computational code [19]. The exchange-correlation energy of the electrons is described according to the local-density approximation (LDA) framework [20]. The Coulomb potential energy caused by electron-ion interaction is described using the ultrasoft scheme, in which the orbits of Gd(*$4f^7$, $5s^2$, $5p^6$, $5d^1$, $6s^2$*), Zr(*$4s^2$, $4p^6$, $4d^2$, $5s^2$*), U(*$5f^3$, $6s^2$, $6p^6$, $6d^1$, $7s^2$*) and O(*$2s^2$, $2p^4$*) are treated as valence electrons. The electronic wave functions are expanded in a plane wave basis set with an energy cutoff of 500 eV. A Monkhorst-Pack mesh with 3×3×3 *k*-points is used for the Brillouin-zone k-points sampling, and the self-consistent convergence of total energy is $2 \times 10^{-6}$ eV/atom. These parameters are sufficient to achieve well-converged total energy and geometrical configurations.

The VCA method is used to model four series of configurations, including $(Gd_{2-y}U_y)Zr_2O_7$, $Gd_2(Zr_{2-y}U_y)O_7$, $(Gd_{2-y}Zr_y)(Zr_{2-y}Gd_y)O_7$ and $(Gd_{2-y}Zr_y)(Zr_{2-y}U_y)O_7$, to study the solution behavior of uranium at the different sites of a $Gd_2Zr_2O_7$ lattice. The VCA in *CASTEP* is capable of correctly describing a number of mixture atoms [21]. The VCA method could be performed well even when it was applied to elements far from each other in the periodic table [22-25]. The atomic sites in a disordered crystal can be described in terms of a hybrid atom, which consists of two element types. The relative concentrations can be set for any number of atoms, and the total concentration must be 100%. Using the VCA method, the uranium atoms can be distributed randomly in the $Gd_2Zr_2O_7$ lattice, and any percentage of uranium content can be introduced to the disordered system.

The full structural relaxations for all of the configurations are performed firstly. During the structure optimizations, the total energy is minimized by varying the lattice constants according to the restriction of the given symmetry. Then, the crystal and electronic structures of all of the series are calculated. As structural-phase changes from pyrochlore to fluorite are observed at higher *y* values [11], *y*=0-0.4 for all of the series in this study. The VCA method is adopted, according to which the uranium atoms occupying the lattice site are described via the index *y* (i.e., the atomic fraction of uranium at the lattice site) to model the solubility of the uranium in the $Gd_2Zr_2O_7$ pyrochlore. The calculation results are statistical average values.

**Table 1.** The lattice constants, internal positional parameters $x_{O48f}$, band gap energies and bond distances between the nearest-neighboring cations and anions of the $Gd_2Zr_2O_7$ pyrochlore.

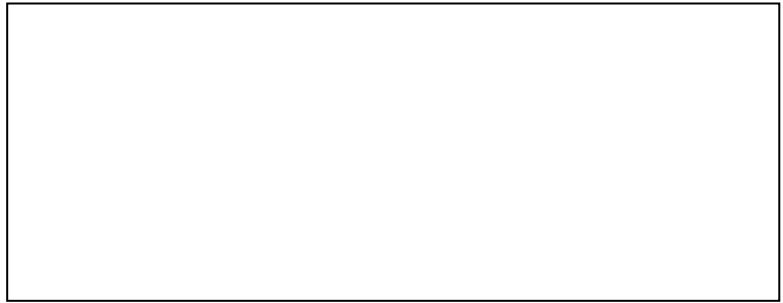

## 3 Results and discussion

Table 1 presents the lattice constants, internal positional parameter $x_{O48f}$ and bond distances between the nearest-neighboring cations and anions of the $Gd_2Zr_2O_7$ pyrochlore. The calculated lattice constants $a$ and $x_{O48f}$ are 10.37 Å and 0.340, respectively, and therefore in agreement with experimental [18,26] and other calculated [9,27-28] results. The bond distances between the nearest-neighboring cations and anions are also in agreement with other results [26]. The tight-binding (TB) [9] and DFT with an energy correction (DFT+U) approach [23] exhibit bigger lattice constants and bond lengths of $Gd_2Zr_2O_7$ pyrochlore. Although the TB approach is based on quantum mechanics, it lacks reliability and transferability due to the parameterization of the electronic Hamiltonian with a finite set of equilibrium structures and properties, and even lacks self-consistency in most cases.

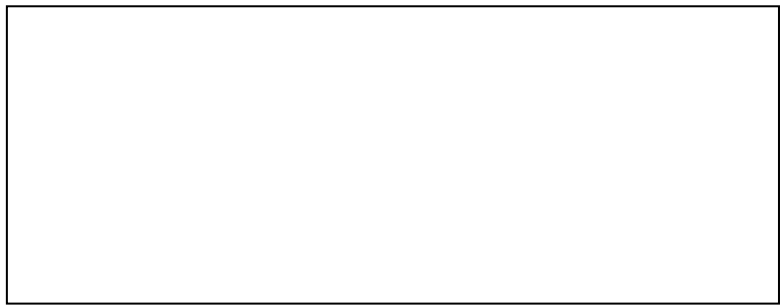

**Fig. 1.** The calculated lattice constants $a$ as a function of uranium content $y$ (in $(Gd_{2-y}U_y)Zr_2O_7$, $Gd_2(Zr_{2-y}U_y)O_7$, and $(Gd_{2-y}Zr_y)(Zr_{2-y}U_y)O_7$) and cation disorder parameter $y$ (in $(Gd_{2-y}Zr_y)(Zr_{2-y}Gd_y)O_7$): (a) the lattice constants $a$ of $(Gd_{2-y}U_y)Zr_2O_7$ and $Gd_2(Zr_{2-y}U_y)O_7$; and (b) the lattice constants $a$ of $(Gd_{2-y}Zr_y)(Zr_{2-y}U_y)O_7$ and $(Gd_{2-y}Zr_y)(Zr_{2-y}Gd_y)O_7$.

The cubic lattice constant and positional parameter of $x_{O48f}$ can completely describe the pyrochlore structure. Fig. 1(a) exhibits the calculated lattice constants $a$ as a function of uranium content $y$ in $(Gd_{2-y}U_y)Zr_2O_7$ and $Gd_2(Zr_{2-y}U_y)O_7$ to clarify the fundamental properties of U-doped gadolinium zirconate pyrochlore. Vegard's law is valid below $y=0.3$, and the lattice constants of $Gd_2(Zr_{2-y}U_y)O_7$ and $(Gd_{2-y}U_y)Zr_2O_7$ are generally expressed as being linearly related to the uranium content $y$. Although the lattice constants of $Gd_2(Zr_{2-y}U_y)O_7$ increase along with the enriched uranium content, the $(Gd_{2-y}U_y)Zr_2O_7$ series shows the opposite tendency.

It has been observed that the increase in the lattice constant of $(Gd_{2-y}U_y)Zr_2O_7$ is related to the difference in ionic radii of the gadolinium and uranium ions, and that the decrease of $Gd_2(Zr_{2-y}U_y)O_7$ is related to the incorporation of the larger uranium ion at the zirconium site [11] The differences between the calculated lattice parameter and available experimental results are all within 1.7%, indicating that the calculation results of the current study are fairly trustworthy. The lattice expansion is obviously expressed along with the degree of cation disorder, as shown in Fig. 1(b). Previous studies have also expressed linear lattice constants related to the indium composition $x$ in $In_xAl_{1-x}N$ and to the cation disorder index $x$ in $(Mg_{1-x}Al_x)[Mg_xAl_{2-x}]O4$ [29-30].

The lattice constants present obvious variations when $y>0.4$ (data not shown). The variation mainly involves the loss of pyrochlore ordering above this concentration level [11]. Therefore, the pyrochlore structure model for high uranium content is not suitable. The results of $y<0.2$ are considered in the following discussion.

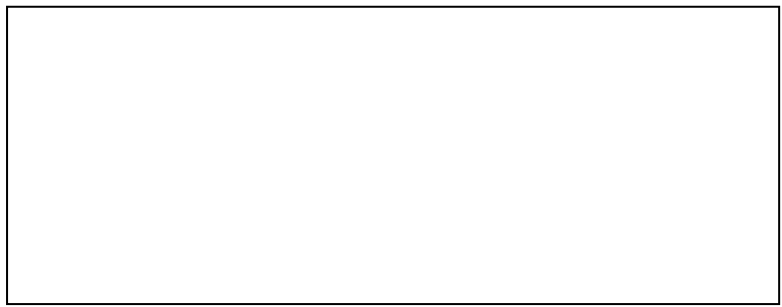

**Fig. 2.** The variations in the optimized bond length as a function of uranium content $y$ (in $(Gd_{2-y}U_y)Zr_2O_7$, $Gd_2(Zr_{2-y}U_y)O_7$ and $(Gd_{2-y}Zr_y)(Zr_{2-y}U_y)O_7$) and cation disorder parameter $y$ (in $(Gd_{2-y}Zr_y)(Zr_{2-y}Gd_y)O_7$) (a) $d_{A-O8b}$: bond length between A-site cation and $O_{8b}$; (b) $d_{A-48f}$: bond length between A-site cation and $O_{48f}$; (c) $d_{B-48f}$: bond length between B-site cation and $O_{48f}$.

Fig. 2 shows the variations in the optimized bond length as a function of uranium content $y$ or cation disorder parameter $y$. Although the bond length between O and its nearest-neighboring cations gradually increases along with increasing uranium content in $Gd_2(Zr_{2-y}U_y)O_7$, the opposite holds true for $(Gd_{2-y}U_y)Zr_2O_7$. No obvious changes in the bond length between the O and B-site cations are observed. The bond lengths $d_{A-O8b}$ and $d_{B-48f}$ obviously increase along with the increasing cation disorder parameter $y$ in $(Gd_{2-y}Zr_y)(Zr_{2-y}U_y)O_7$. The change in tendency of the lattice parameters (see Fig. 1) is similar to the change in tendency of $d_{A-O8b}$ (the bond length between the A-site cation and $O_{8b}$) and $d_{B-O48f}$ (the bond length between the B-site cation and $O_{48f}$), exhibiting a certain regularity and a preferable relativity. The lattice constant of the pyrochlore is determined by $d_{A-O8b}$ and $d_{B-O48f}$, which are small bond lengths exhibit strong interactions between the related atoms.

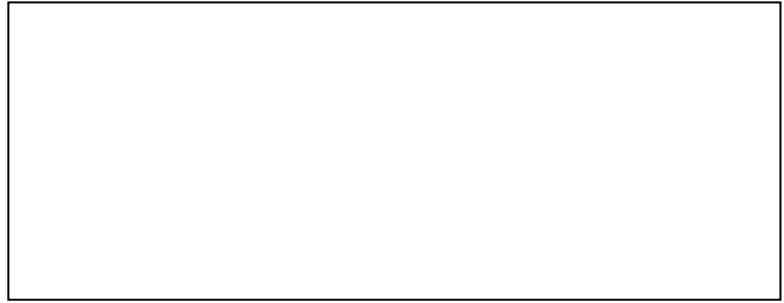

**Fig. 3.** Electron density difference of (a) $Gd_2Zr_2O_7$; (b) $Gd_2(Zr_{2-y}U_y)O_7$; (c) $(Gd_{2-y}U_y)Zr_2O_7$; (d) $(Gd_{2-y}Zr_y)(Zr_{2-y}Gd_y)O_7$; and (e) $(Gd_{2-y}Zr_y)(Zr_{2-y}U_y)O_7$.

The electron density difference is typically the difference between an assumed standard or model electron densities and the actual observed or DFT-computed electron density. The difference provides information related to electron transfer and distribution after bonding. The electron density differences of $Gd_2Zr_2O_7$, $Gd_2(Zr_{2-y}U_y)O_7$, $(Gd_{2-y}U_y)Zr_2O_7$, $(Gd_{2-y}Zr_y)(Zr_{2-y}Gd_y)O_7$ and $(Gd_{2-y}Zr_y)(Zr_{2-y}U_y)O_7$ are shown in Fig. 3, where the bright (red) and darker (blue) colored areas indicate positive and negative electron transfers, respectively.

In general, there exists a covalent interaction between the B-site cations and nearest-neighboring O, and an ionic interaction between the A-site cations and nearest-neighboring O. There are no obvious electronegativity changes in $(Gd_{2-y}U_y)Zr_2O_7$ and $Gd_2(Zr_{2-y}U_y)O_7$. The changes in the lattice constants of $(Gd_{2-y}U_y)Zr_2O_7$ and $Gd_2(Zr_{2-y}U_y)O_7$ are related to the difference in the ionic radii change after doping. Furthermore, the lattice-constant changes of $(Gd_{2-y}Zr_y)(Zr_{2-y}Gd_y)O_7$ and $(Gd_{2-y}Zr_y)(Zr_{2-y}U_y)O_7$ are related to the electronegativity change of the cation sites arising from the atom swapping or doping.

There is a high level of Coulomb repulsion between two high-electronegativity atoms. As shown in Fig. 3(d), the A-site electronegativity in the $(Gd_{2-y}Zr_y)(Zr_{2-y}Gd_y)O_7$ lattice obviously increases (more red areas) due to cation disorder. The Coulomb repulsions between the A-site atoms and $O_{8b}$ increase, and the ionic bond $d_{A-O8b}$ increases while the coulomb repulsions are dominant. The covalent bond plays a dominant role in the bond $d_{B-O48f}$, which increases due to the introduction of atoms with larger ionic radii in the B sites.

The lattice pyrochlore constant is determined by the small bond lengths $d_{A-O8b}$ and $d_{B-O48f}$. Therefore, the lattice parameter of $(Gd_{2-y}Zr_y)(Zr_{2-y}Gd_y)O_7$ obviously increases with the increase in cation disorder, and there are fewer lattice parameter increases in the $(Gd_{2-y}Zr_y)(Zr_{2-y}U_y)O_7$ case due to the lesser increase in electronegativity in the B-site atoms.

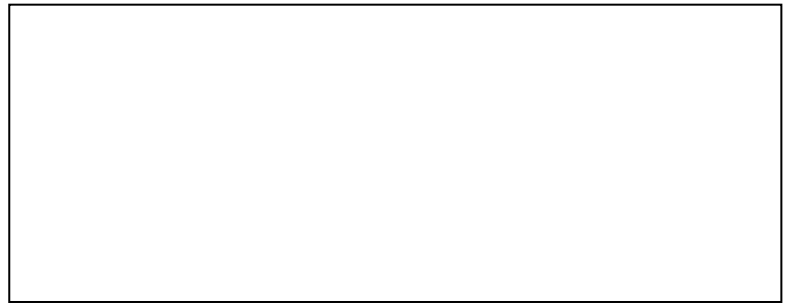

**Fig. 4.** Total energy as a function of *y* (uranium content and cation disorder parameter).

The total energy of $(Gd_{2-y}U_y)Zr_2O_7$ and $(Gd_{2-y}Zr_y)(Zr_{2-y}U_y)O_7$ gradually increases with the uranium content $y$ (see Fig. 4). With the increase of cation disorder $y$, the total energy of $(Gd_{2-y}Zr_y)(Zr_{2-y}Gd_y)O_7$ increases linearly, suggesting the instability of the higher cation disorder. This is comparable with the total energy difference between $(Gd_{2-y}U_y)Zr_2O_7$ and $(Gd_{2-y}Zr_y)(Zr_{2-y}U_y)O_7$ of the same uranium content y, as the two structures have the same numbers and kinds of atoms. $(Gd_{2-y}Zr_y)(Zr_{2-y}U_y)O_7$ structures have a smaller total energy and are more stable than $(Gd_{2-y}U_y)Zr_2O_7$ structures when the $y$ values are the same.

Synchrotron XRD, XANES and positron annihilation lifetime spectroscopy studies have suggested that the uranium cation in doped gadolinium zirconates ($Gd_{2-x}U_xZr_2O_7$, $x$=0.1 and 0.2) is largely located at the pyrochlore B site instead of the targeted A site [18]. Based on the total energy calculation, uranium is a more preferable substitute for the B-site gadolinium atoms in cation-disordered $Gd_2Zr_2O_7$ than for the A-site gadolinium atoms in ordered $Gd_2Zr_2O_7$ due to the lower total energy of the former. Finally, there is little total energy variation in $(Gd_{2-y}Zr_y)(Zr_{2-y}U_y)O_7$ due to the small differences in the atomic energy and ionic radius between zirconium and uranium [31].

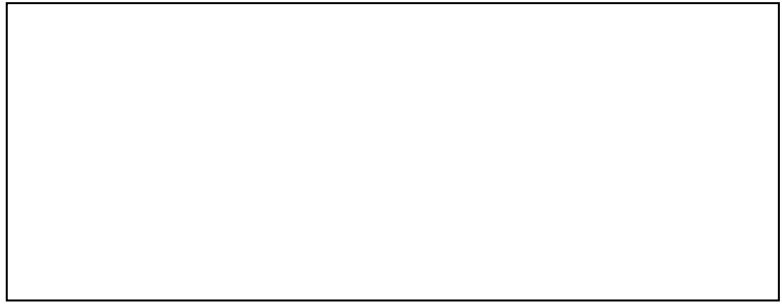

**Fig. 5.** The $O_{48f}$ positional parameter $x$ ($x_{O48f}$) as a function of $y$ (uranium content and cation disorder parameter).

All of the atoms in an ideal pyrochlore except $48f$ ($O_{48f}$) are located at special positions. Thus, the structure is completely described by the lattice parameter $a$ and the fractional coordinate $x$ of $O_{48f}$ [32]. The $O_{48f}$ positional parameter $x$ ($x_{O48f}$) is an important indicator value of the degree of disorder and resistance to irradiation-induced amorphization of a pyrochlore [33]. As shown in Fig. 5, the $x_{O48f}$ values of U-doped pyrochlore do not obviously change with an increase in uranium content. This suggests that the gadolinium zirconate pyrochlore is a good uranium immobilization matrix that can maintain its radiation tolerance in cases of low uranium content doping. The $x_{O48f}$ values of the cation disorder series $(Gd_{2-y}Zr_y)(Zr_{2-y}U_y)O_7$ obviously increase when $y>0.1$.

Lian *et al*. [34] observed that the radiation resistance of pyrochlore was closely related to the cation ionic radius ratio and *x* positional parameter $x_{O48f}$. A decrease in the average ionic radius ratio of the A and B sites ($r_A/r_B$) generally leads to a decrease in the critical temperature for amorphization and a resistance to radiation damage. Furthermore, the radiation resistance of $Gd_2(Ti_{1-x}Zr_x)_2O_7$ increases as the concentration of zirconium increases [35]. The high radiation resistance of $Gd_2Zr_2O_7$ is caused by not only the high neutron absorption cross-section of gadolinium, but also the transformation of the pyrochlore structure into a defect-fluorite structure, with the disordering of cation antisite defects (A and B cations that exchange places with each other) coupled with the disordering of oxygen Frenkel pairs (a *48f* oxygen moved to an adjacent empty *8a* site, resulting in a vacant *48f* oxygen site and an occupied *8a* oxygen site).

Gregg *et al*. [18] suggested that cation disordering is important to an increased ionic conductivity and an increased resistance to radiation-induced amorphization in pyrochlores. Pyrochlores have the general formula $A_2B_2O_7$, where A is usually occupied by the larger cation and B is occupied by the smaller cation. The stability of the pyrochlore structure is governed by the ionic radii ratio of the A and B cations ($r_A/r_B$) [9], which is affected by disorder, ionic value and the substitution site of the doping atom. The high radiation resistance of the $Gd_2Zr_2O_7$ pyrochlore can be attributed to the similar iron radii of the A- and B-site atoms, which easily change into highly radiation-resistant fluorite structures [3]. It has been a consistent conclusion that the radii ratio of the A and B sites ($r_A/r_B$) decreases with the increase in *y* in $(Gd_{2-y}Zr_y)(Zr_{2-y}Gd_y)O_7$. Furthermore, the positional parameter $x_{O48f}$ increases, showing a higher radiation resistance capability.

## 4 Conclusions

In this study, the VCA method is used to calculate the total energy and structural properties of U-doped $Gd_2Zr_2O_7$ pyrochlores (i.e., both ordered and cation-disordered structures). The $Gd_2Zr_2O_7$ maintains its pyrochlore structure at low uranium dopant levels and the lattice constants of $Gd_2(Zr_{2-y}U_y)O_7$ and $(Gd_{2-y}U_y)Zr_2O_7$ are generally expressed as a linear Vegard's relation to the uranium content $y$. As the uranium content is enriched, the lattice constants of $Gd_2(Zr_{2-y}U_y)O_7$ increase. However, the $(Gd_{2-y}U_y)Zr_2O_7$ series exhibits the opposite tendency. The results are in agreement with experiments.

The lattice-constant changes of $(Gd_{2-y}U_y)Zr_2O_7$ and $Gd_2(Zr_{2-y}U_y)O_7$ are related to the difference in the ionic radii change after doping. The lattice-constant changes of $(Gd_{2-y}Zr_y)(Zr_{2-y}Gd_y)O_7$ and $(Gd_{2-y}Zr_y)(Zr_{2-y}U_y)O_7$ are related to the change in electronegativity of the cation sites arising from the swapping or doping of atoms. In the case of $(Gd_{2-y}Zr_y)(Zr_{2-y}Gd_y)O_7$, the Coulomb repulsions between the A-site atoms and $O_{8b}$-site increase, and the ionic bond $d_{A-O8b}$ increases while the Coulomb repulsions are dominant. The covalent bond plays a dominant role in the bond $d_{B-O48f}$, which increases due to the introduction of atoms with larger ionic radii. The lattice constant of pyrochlore is determined by the small bond lengths of $d_{A-O8b}$ and $d_{B-O48f}$. Therefore, the lattice parameter obviously increases along with an increase in cation disorder in $(Gd_{2-y}Zr_y)(Zr_{2-y}Gd_y)O_7$, and there is less of a lattice parameter increase in $(Gd_{2-y}Zr_y)(Zr_{2-y}U_y)O_7$ due to the lower electronegativity increase at the B site. The radii ratio of the A and B sites ($r_A/r_B$) decreases as the cation disorder $y$ in $(Gd_{2-y}Zr_y)(Zr_{2-y}Gd_y)O_7$ increases. Furthermore, the positional parameter $x_{O48f}$ increases, indicating a higher radiation resistance capability. Finally, uranium is a preferable substitute for the B-site gadolinium atoms in cation-disordered $Gd_2Zr_2O_7$ over the A-site gadolinium atoms in ordered $Gd_2Zr_2O_7$ due to the lower total energy of the former.

**Table 1** The lattice constants, internal positional parameters $x_{O48f}$, band gap energies and bond distances between the nearest-neighboring cations and anions of the $Gd_2Zr_2O_7$ pyrochlore.

|  | $a$ (Å) | $X$ | $d_{Gd-O8b}$ (Å) | $d_{Gd-O48f}$ (Å) | $d_{Zr-O48f}$ (Å) |
|---|---|---|---|---|---|
| LDA | 10.37 | 0.340 | 2.246 | 2.470 | 2.060 |
| Exp. | 10.54 [19] | 0.344 [19] |  | 2.483 [19] | 2.110 [19] |
|  | 10.53 [15] |  |  |  |  |
| Other calc. | 10.535 [9] | 0.374 [9] | 2.281 [9] | 2.462 [9] | 2.125 [9] |
|  | 10.66 [20] | 0.339 [20] | 2.307 [20] | 2.548 [20] | 2.110 [20] |
|  | 10.76 [21] |  |  |  |  |

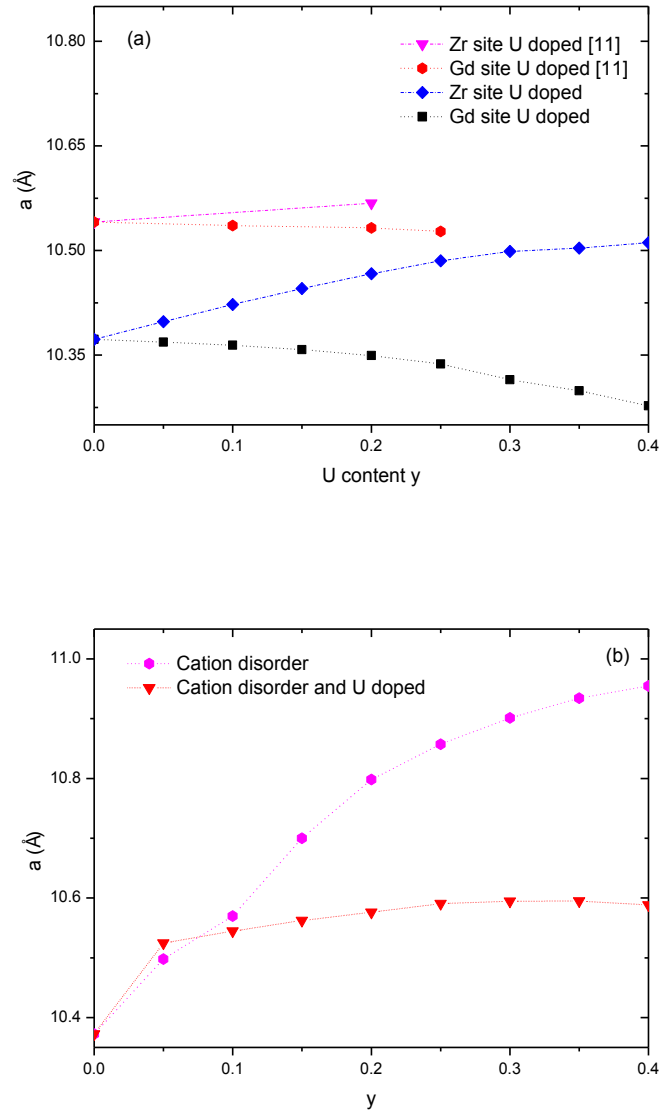

**Fig. 1.** The calculated lattice constants *a* as a function of uranium content *y* (in $(Gd_{2-y}U_y)Zr_2O_7$, $Gd_2(Zr_{2-y}U_y)O_7$, and $(Gd_{2-y}Zr_y)(Zr_{2-y}U_y)O_7$) and cation disorder parameter *y* (in $(Gd_{2-y}Zr_y)(Zr_{2-y}Gd_y)O_7$): (a) the lattice constants *a* of $(Gd_{2-y}U_y)Zr_2O_7$ and $Gd_2(Zr_{2-y}U_y)O_7$; and (b) the lattice constants *a* of $(Gd_{2-y}Zr_y)(Zr_{2-y}U_y)O_7$ and $(Gd_{2-y}Zr_y)(Zr_{2-y}Gd_y)O_7$.

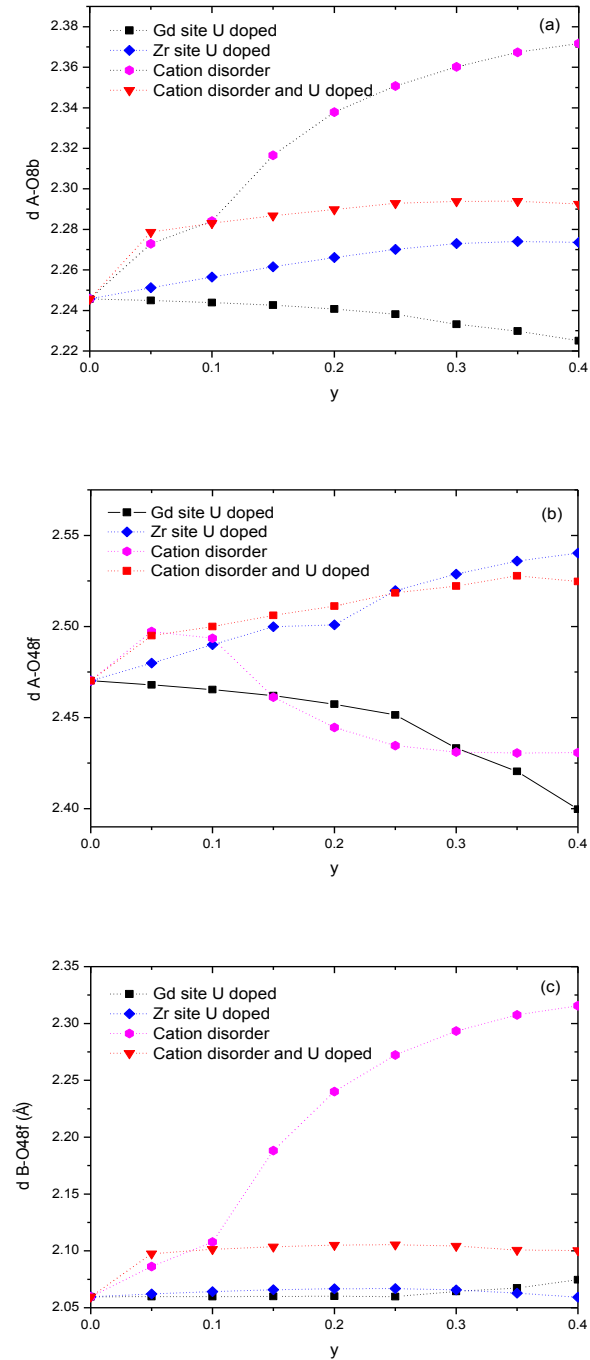

**Fig. 2.** The variations in the optimized bond length as a function of uranium content $y$ (in $(Gd_{2-y}U_y)Zr_2O_7$, $Gd_2(Zr_{2-y}U_y)O_7$ and $(Gd_{2-y}Zr_y)(Zr_{2-y}U_y)O_7$) and cation disorder parameter $y$ (in $(Gd_{2-y}Zr_y)(Zr_{2-y}Gd_y)O_7$) (a) $d_{A-O8b}$: bond length between A-site cation and $O_{8b}$; (b) $d_{A-48f}$: bond length between A-site cation and $O_{48f}$; (c) $d_{B-48f}$: bond length between B-site cation and $O_{48f}$.

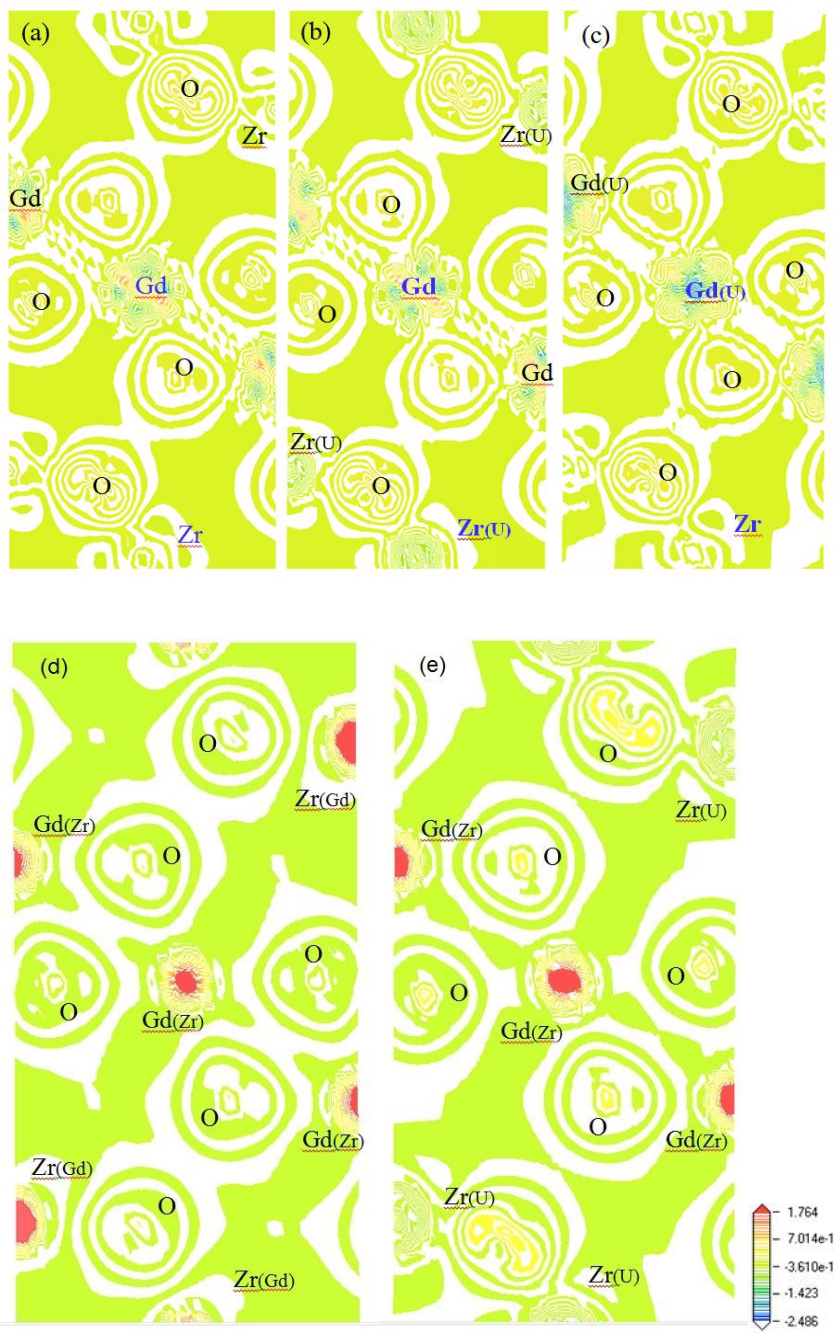

**Fig. 3.** Electron density difference of (a) $Gd_2Zr_2O_7$; (b) $Gd_2(Zr_{2-y}U_y)O_7$; (c) $(Gd_{2-y}U_y)Zr_2O_7$; (d) $(Gd_{2-y}Zr_y)(Zr_{2-y}Gd_y)O_7$; and (e) $(Gd_{2-y}Zr_y)(Zr_{2-y}U_y)O_7$.

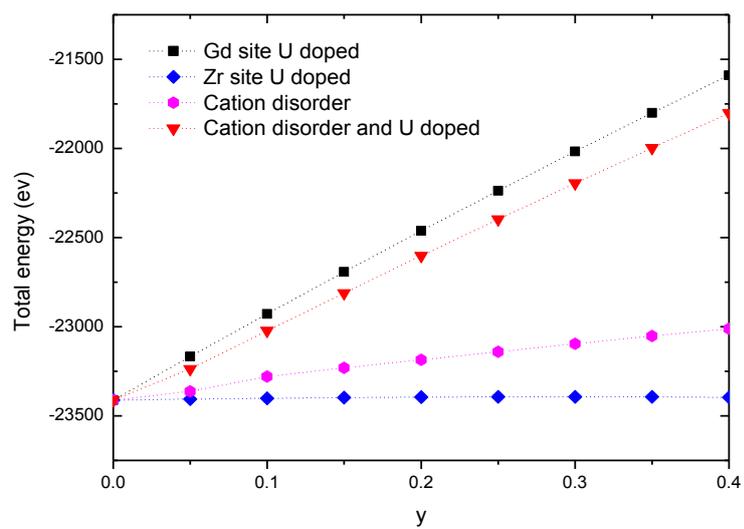

**Fig. 4.** Total energy as a function of *y* (uranium content and cation disorder parameter).

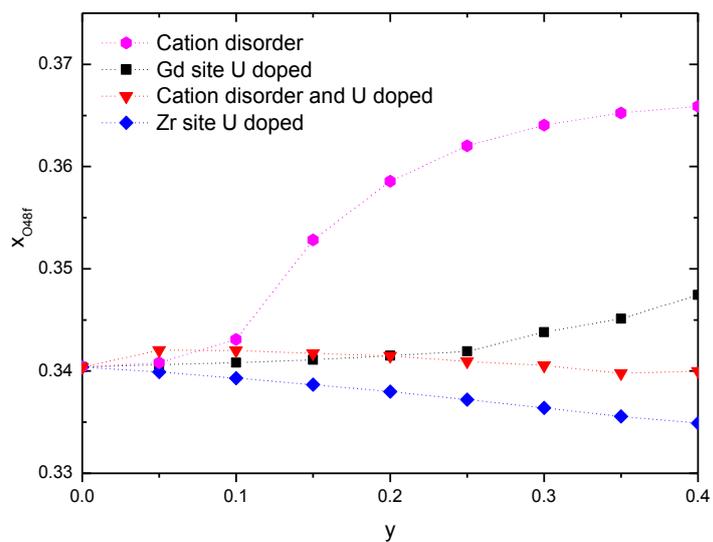

**Fig. 5.** The $O_{48f}$ positional parameter $x$ ($x_{O48f}$) as a function of $y$ (uranium content and cation disorder parameter).